\documentclass{aa}
\usepackage[varg]{txfonts}

\title{De-blending Deep \textit{Herschel}\thanks{{\it Herschel} is an ESA space observatory with science instruments provided by European-led Principal Investigator consortia and with important participation from NASA.} Surveys:\\A Multi-wavelength Approach}

\author{W.~J.~Pearson\inst{\ref{inst:SRON}, \ref{inst:Kapteyn}}, L.~Wang\inst{\ref{inst:SRON}, \ref{inst:Kapteyn}}, F.~F.~S.~van~der~Tak\inst{\ref{inst:SRON}, \ref{inst:Kapteyn}}, P.~D.~Hurley\inst{\ref{inst:Sussex}}, D.~Burgarella\inst{\ref{inst:LAM}}, S.~J.~Oliver\inst{\ref{inst:Sussex}}}

\authorrunning{W.~J.~Pearson et al.}
\institute{SRON Netherlands Institute for Space Research, Landleven 12, 9747 AD, Groningen, The Netherlands\label{inst:SRON}\\\email{w.j.pearson@sron.nl}
\and Kapteyn Astronomical Institute, University of Groningen, Postbus 800, 9700 AV Groningen, The Netherlands\label{inst:Kapteyn}
\and Astronomy Centre, Department of Physics and Astronomy, University of Sussex, Falmer, Brighton BN1 9QH, UK\label{inst:Sussex}
\and Aix-Marseille Universit\'{e}, CNRS, LAM (Laboratorie d'Astrophysique de Marseille) UMR 7326, 13388 Marseille, France\label{inst:LAM}
}

\keywords{Galaxies: statistics -- Infrared: galaxies}

\usepackage{natbib}
\bibpunct{(}{)}{;}{a}{}{,}

\date{Received 21 November 2016 /
Accepted 06 April 2017}
\abstract{}
{Cosmological surveys in the far infrared are known to suffer from confusion. The Bayesian de-blending tool, XID+, currently provides one of the best ways to de-confuse deep \textit{Herschel} SPIRE images, using a flat flux density prior. This work is to demonstrate that existing multi-wavelength data sets can be exploited to improve XID+ by providing an informed prior, resulting in more accurate and precise extracted flux densities.}
{Photometric data for galaxies in the COSMOS field were used to constrain spectral energy distributions (SEDs) using the fitting tool CIGALE. These SEDs were used to create Gaussian prior estimates in the SPIRE bands for XID+. The multi-wavelength photometry and the extracted SPIRE flux densities were run through CIGALE again to allow us to compare the performance of the two priors. Inferred ALMA flux densities (F$^{\mathrm{infer}}_{\mathrm{ALMA}}$), at 870 $\mu$m and 1250 $\mu$m, from the best fitting SEDs from the second CIGALE run were compared with measured ALMA flux densities (F$^{\mathrm{meas}}_{\mathrm{ALMA}}$) as an independent performance validation. Similar validations were conducted with the SED modelling and fitting tool MAGPHYS and modified black body functions to test for model dependency.}
{We demonstrate a clear improvement in agreement between the flux densities extracted with XID+ and existing data at other wavelengths when using the new informed Gaussian prior over the original uninformed prior. The residuals between F$^{\mathrm{meas}}_{\mathrm{ALMA}}$ and F$^{\mathrm{infer}}_{\mathrm{ALMA}}$ were calculated. For the Gaussian prior, these residuals, expressed as a multiple of the ALMA error ($\sigma$), have a smaller standard deviation, 7.95~$\sigma$ for the Gaussian prior compared to 12.21~$\sigma$ for the flat prior, reduced mean, 1.83~$\sigma$ compared to 3.44~$\sigma$, and have reduced skew to positive values, 7.97 compared to 11.50. These results were determined to not be significantly model dependent. This results in statistically more reliable SPIRE flux densities and hence statistically more reliable infrared luminosity estimates.}
{}

\begin{document}
\maketitle
\addtocounter{footnote}{4}
\section{Introduction} \label{sec:intro}
Infrared (IR) radiation makes up approximately half of the total extragalactic emission that we observe \citep[e.g.][]{2006AA...451..417D, 2013A&A...554A..70B}. As a result, it is important to observe at these wavelengths to gain a better understanding of our Universe. ESA's \textit{Herschel Space Observatory's} \citep{2010AA...518L...1P} Spectral and Photometric Imaging Receiver \citep[SPIRE;][]{2010AA...518L...3G} cosmological surveys probe the far IR but are known to suffer from source confusion \citep[e.g.][]{2010A&A...518L...5N, 2012MNRAS.424.1614O}, with object separation in the COSMOS2015 catalogue \citep{2016ApJS..224...24L} $\lesssim$~8$^{\prime\prime}$ and SPIRE's smallest beam size of 18$^{\prime\prime}$ \citep{2010AA...518L...3G}. There have been a number of tools created to de-blend SPIRE images, such as DESPHOT \citep{2010MNRAS.409...48R,2012MNRAS.419.2758R,2014MNRAS.444.2870W} or T-PHOT \citep{2015AA...582A..15M}. These tools mostly use maximum likelihood estimation to generate flux density estimates using galaxy positions extracted from a shorter wavelength image.

More recently, the probabilistic, Bayesian de-blender XID+ \citep{2017MNRAS.464..885H} has been developed, overcoming the main weaknesses of DESPHOT: its tendency to assign all the flux density to one source when many sources are within a single beam as well as the poor estimation of variance and co-variance of sources. This is achieved by exploring the full posterior distribution (see Sect. \ref{subsubsec:xidp}) using the Bayesian statistical inference tool Stan \citep{JSSv076i01}. As a result of posterior exploration, XID+ produces much better flux density precision with more realistic associated uncertainties than DESPHOT across all three SPIRE bands \citep{2017MNRAS.464..885H}. However, the current, publicly available XID+ only uses a flat and uninformed flux density prior.

The aim of this paper is to show how the performance of XID+ can be advanced by exploiting the large amount of multi-wavelength data available in the popular COSMOS field \citep{2007ApJS..172....1S} as part of the continuing development of XID+, although the technique can be applied in any deep field with multi-wavelength data available. Throughout this paper, Wilkinson Microwave Anisotropy Probe year 7 cosmology \citep{2011ApJS..192...18K, 2011ApJS..192...16L} is followed: $\Omega_{\mathrm{M}}$ = 0.27, $\Omega_{\Lambda}$ = 0.73 and H$_{0}$ = 70.4 km\,s$^{-1}$\,Mpc$^{-1}$.

\section{Data} \label{sec:data}
For this work, a large multi-wavelength data set from a field covered by the \textit{Herschel} Multi-tiered Extragalactic Survey \citep[HerMES;][]{2012MNRAS.424.1614O} was required. The multi-wavelength data is needed to generate flux density priors for de-blending the SPIRE images from HerMES. A secondary data set is also required for validation of the extracted flux densities. For this we choose data from the Atacama Large Millimeter/submillimeter Array (ALMA) due to its higher resolution coupled with the fact that the 850~$\mu$m and 1100~$\mu$m bands are dominated by the same emission as the SPIRE bands \citep[e.g.][]{2014ApJ...783...84S, 2016ApJ...820...83S}.

Thermal emission dominates at wavelengths of approximately 50~$\mu$m or more and contains the Rayleigh-Jeans (RJ) tail from approximately 100~$\mu$m to 1100~$\mu$m in the rest frame \citep[e.g.][]{2003ARAA..41..241D, 2007ApJ...657..810D}. Thus, at least one SPIRE band lies within the RJ tail up to redshifts of 4 and all three SPIRE bands will be dominated by thermal emission at all redshifts studied here. Similarly, the 870~$\mu$m ALMA band will remain in the RJ tail for the entirety of this study while the 1250~$\mu$m ALMA band enters the RJ region at redshifts of 0.14, hence why it is chosen for high redshift objects only. Evidently, the SPIRE bands and the two ALMA bands chosen for this study are both dominated by the same physical processes, certainly up to redshifts of at least 4, making ALMA data a good choice to validate the SPIRE data.

The COSMOS field was chosen due to the prevalence of multi-wavelength data as well as the SPIRE images within COSMOS being considered to be reasonably homogenous and of high quality (see Appendix \ref{app:noise}). The COSMOS2015 catalogue \citep{2016ApJS..224...24L} was used as an ancillary data set as it contains data on over 1.2 million objects in over 30 wavelength bands. For this work, only bands from the ultra violet to the mid IR were used. The \textit{Subaru} narrow bands were omitted as similar wavelengths are covered by the \textit{Subaru} intermediate bands\footnote{We note that the \textit{Subaru} narrow bands are used in the photometric redshift derivation in the COSMOS2015 catalogue.}. The \textit{Spitzer} Multiband Imaging Photometer \citep{2004ApJS..154...25R} 24~$\mu$m data were also not used as they also noticeably suffer from confusion, although not to the same extent as the SPIRE data \citep[e.g.][]{2004ApJS..154...93D, 2004ApJS..154...25R}. It also lies in a region of the spectrum that is more complicated to model, with contributions from active galactic nuclei (AGN), thermal dust emission and non-thermal polycyclic aromatic hydrocarbons. The $3^{\prime\prime}$ aperture data were used, where available, to match the \textit{Spitzer} Infrared Array Camera data.

The latest SPIRE maps for the COSMOS field, Data Release 4 from HerMES \citep{2012MNRAS.424.1614O}, from the \textit{Herschel} Database in Marseille\footnote{\url{http://hedam.lam.fr}} were used for the SPIRE extraction. These maps have a beam size of 18.1$^{\prime\prime}$, 25.2$^{\prime\prime}$ and 36.6$^{\prime\prime}$ and 5$\sigma$ confusion limits of 24.0, 27.5 and 30.5 mJy for the 250~$\mu$m, 350~$\mu$m and 500~$\mu$m bands respectively \citep{2010AA...518L...3G, 2010A&A...518L...5N}.

The ALMA archive\footnote{\url{https://almascience.nrao.edu/alma-data/archive}} was searched for objects within the COSMOS field with positive flux densities at 870~$\mu$m, or 1250~$\mu$m. A total of 214 objects were found that met these criteria: 192 objects with 870~$\mu$m data at $z < 3$ from \citet{2014ApJ...783...84S, 2016ApJ...820...83S} and 22 objects with 1250~$\mu$m data at $z > 3$ from \citet{2016ApJ...820...83S}. In the original studies, these objects were selected as they have stellar masses of approximately 10$^{11}$~M$_{\odot}$. Of the 214 objects, 43 (20\%) have a signal-to-noise ratio of less than 2$\sigma$, all of which are from \citet{2014ApJ...783...84S} due to the higher values of 1$\sigma$ rms noise. The reader is referred to \citet{2014ApJ...783...84S} and \citet{2016ApJ...820...83S} for discussion on the signal-to-noise of the ALMA observations.

The ALMA data were matched to the COSMOS2015 sources. With a positional accuracy of 0.15$^{\prime\prime}$ for the COSMOS2015 catalogue \citep{2016ApJS..224...24L} and a pointing accuracy of 0.6$^{\prime\prime}$ for ALMA, matching within 1$^{\prime\prime}$ of the ALMA sources was deemed adequate. For one ALMA object there was more than one match to the COSMOS2015 catalogue, so the closest object was used. Once matched, the photometric redshifts from COSMOS2015 were used.

\section{Methodology} \label{sec:method}
\subsection{Tools} \label{subsec:tool}
\subsubsection{XID+} \label{subsubsec:xidp}
XID+\footnote{\url{https://github.com/H-E-L-P/XID_plus}} \citep{2017MNRAS.464..885H} is a probabilistic de-blending tool created to extract source flux densities from photometry maps that suffer from source confusion. This is achieved by using Bayesian inference to explore the posterior distribution function. Once converged, the flux density is reported along with the upper and lower 1 sigma uncertainties. In the original version, XID+ uses a flat prior in parameter space, between zero and the brightest value in the map, along with the source positions on the sky. This work introduces a more informed Gaussian prior, again truncated between zero and the brightest value in the map. The mean and sigma for these Gaussian priors are generated by using CIGALE spectral energy distribution (SED) models to estimate the flux densities for the mean and using twice the error on these estimates as the sigma, to be conservative (see also Sect. \ref{subsubsec:cigale} and \ref{subsec:spire}).

To allow parallelisation, which reduces the time XID+ takes to de-blend the map, the map is split up into tiles based on the Hierarchical Equal Area isoLatitude Pixelization of a sphere system \citep[HEALPix;][]{2005ApJ...622..759G} using order 11, which corresponds to an area of 2.95 arcmin$^{2}$ per tile. Order 11 was chosen as it is a compromise between the number of objects in a tile, more objects means a more reliable flux density extraction, and the time it takes a tile to run, here it was required that a tile had a run time of less than one week. XID+ was run using 4 cores per tile on the 162 node Peregrine high performance computing cluster (HPC) at the University of Groningen.

\subsubsection{CIGALE} \label{subsubsec:cigale}
Code Investigating GALaxy Emission\footnote{\url{http://cigale.lam.fr/}} \citep[CIGALE; ][]{2009AA...507.1793N} is a SED modelling and fitting tool with an improved fitting procedure by \citet{2011ApJ...740...22S}. Here, the Python version 0.9.0 is used (Boquien et al. 2017, in prep; Burgarella et al. 2017, in prep) to generate SEDs and fit them to the data from COSMOS2015 to estimate the SPIRE 250~$\mu$m, 350~$\mu$m, and 500~$\mu$m flux densities. CIGALE was run using one node of the Peregrine HPC.

CIGALE models are based around three main components: stars, dust and AGN. The SEDs generated by CIGALE used our choice of a double exponentially declining star formation history (SFH), \citet{2003MNRAS.344.1000B} stellar emission, \citet{2003PASP..115..763C} initial mass function, \citet{2000ApJ...533..682C} dust attenuation, the \citet{2014ApJ...780..172D} update of the \citet{2007ApJ...657..810D} IR dust emission and \citet{2006MNRAS.366..767F} AGN models. A list of parameters used, where they differ from the default values, can be found in Appendix \ref{app:parameters}, along with a brief justification.

\subsection{Extracting the SPIRE Flux Densities}\label{subsec:spire}
CIGALE and XID+ were used to extract the flux densities in the SPIRE bands. The ALMA data were not used in this process. The pipeline begins by using the multi-wavelength data from COSMOS2015 to generate estimates for the SPIRE flux densities and uncertainties using CIGALE. Approximately 1.3\% of the objects in the COSMOS2015 catalogue have no known redshift. These objects were removed from the CIGALE run and assigned an arbitrary flux density, of 7~mJy, along with an artificially large error to produce a functionally flat prior. A further 2.9\% of the COSMOS2015 catalogue are classified as non-galaxies. These objects can be run through CIGALE but care must be taken; these objects will be assigned flux density estimates with the assumption that they have the SED of a galaxy. To compensate for this, the errors were artificially inflated to create an almost flat prior. The objects with no redshift or classed as non-galaxies cannot simply be discarded as they will likely appear in the SPIRE images. For the remaining objects in the COSMOS2015 catalogue with redshifts and classed as galaxies, the errors were multiplied by two to prevent over constraint in XID+ (see also Sec. \ref{subsec:discuss:performance}).

The SPIRE estimates from CIGALE, along with the added data for objects without redshifts, were used as the means in the priors for XID+ while the expanded errors were used as the standard deviations. XID+ was then run on the SPIRE images. As the priors for the objects that are not galaxies, or have no redshift, are effectively flat, the flux for these objects is free to change such that it will not interfere with the fitting of the more constrained sources.

For the flat prior flux densities, the tiles of interest were run through XID+ with a flat flux density prior. These flat prior flux densities were used to compare to the informed prior flux densities.

\section{Results}\label{sec:discuss}
\subsection{CIGALE and XID+ performance}\label{subsec:discuss:performance}
To check if the CIGALE predictions for the SPIRE sources are reasonable, the flux densities for the 100 brightest sources at 250~$\mu$m from the COSMOS2015 catalogue, that also have detections at 350 and 500~$\mu$m, were compared to the predicted flux densities from CIGALE. Blind extraction has been shown to overestimate 250~$\mu$m flux densities by up to 150\%, even for bright sources above the 5$\sigma$ confusion limit \citep{2016MNRAS.460.1119S}. Thus, we expect the flux densities from CIGALE to be systematically lower than those in COSMOS2015. We note, however, that the COSMOS2015 catalogue has SPIRE data extracted using the previous generation de-blending tool DESPHOT, not blind extraction, so the over estimation will not be the same but a comparison is still valid (Scudder 2017 priv. comm.). It is also possible for the CIGALE predictions to under predict for some sources but this is accounted for in our increased error (see below).

Figure \ref{fig:bright} shows that the CIGALE SPIRE predictions are indeed lower than the COSMOS2015 flux densities: the blue data points for the 250~$\mu$m sources fall below the magenta one-to-one line. The cyan line is the locus where the CIGALE flux density is 40\% the COSMOS2015 flux density: the fraction that is expected if the catalogue results are over estimated by 150\%. This line approximately splits the 250~$\mu$m data in half, with 52 objects below the line and 48 above, which would be expected if the catalogue results are 150\% too high. However, there is a large scatter. The 350~$\mu$m and 500~$\mu$m bands are less well split, with 59 and 68 objects below the cyan line respectively. With the lower resolution of the 350~$\mu$m and 500~$\mu$m bands, it is likely that multiplicity will have a greater impact so it is not surprising that the longer wavelengths have more objects with predicted fluxes below the cyan line.

\begin{figure}
	\centering
	\includegraphics[width=0.43\textwidth]{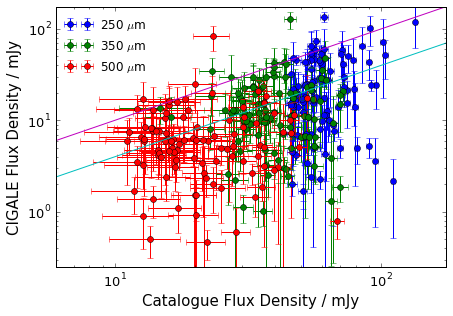}
	\caption{Scatter plot of the predicted CIGALE flux densities against the flux densities from the COSMOS2015 catalogue for the 250$\mu$m (blue points), 350~$\mu$m (green points) and 500~$\mu$m (red points). The errors for the catalogue flux are instrumental noise plus confusion noise. The 1-to-1 line (magenta) and $y = 0.4x$ (cyan) illustrate that the data better fits the idea that bright sources in catalogues are over estimated.}
	\label{fig:bright}
\end{figure}

We can also calculate how many of the 100 bright objects have CIGALE flux densities consistent with 40\% of the catalogue flux densities within CIGALE errors. This is 39, 44 and 45 for the 250~$\mu$m, 350~$\mu$m and 500~$\mu$m bands respectively. If the error is doubled the number of objects increases to 65, 70 and 64 for the 250~$\mu$m, 350~$\mu$m and 500~$\mu$m bands and is consistent with what would be expected for a 1$\sigma$ uncertainty. Therefore, we decided to expand the errors from CIGALE by a factor of two.

\begin{figure}
	\centering
	\includegraphics[width=0.48\textwidth]{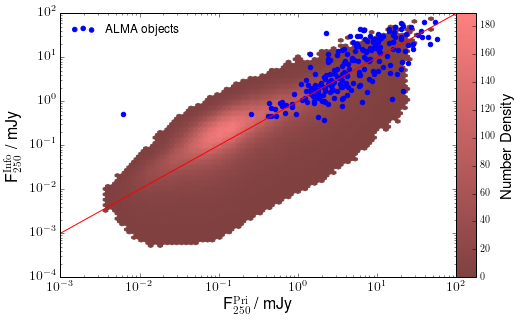}
	\caption{Plot of the extracted 250~$\mu$m flux densities using the informed Gaussian (F$_{250}^{\mathrm{Info}}$) against the 250~$\mu$m flux density priors from CIGALE (F$_{250}^{\mathrm{Pri}}$) for the ALMA selected sources (blue). The fPDF of all objects extracted from the same tiles as the ALMA sources are also shown and presented from high (light red) to low (dark red) density. The red line is the one-to-one line.}
	\label{fig:prior}
\end{figure}

To check that the results from XID+ with an informed prior are not just the prior itself, Fig. \ref{fig:prior} shows a plot of the extracted 250~$\mu$m flux densities using the informed Gaussian against the 250~$\mu$m flux density priors from CIGALE for both the ALMA selected objects (dark blue) as well as the full probability density function derived from the posterior samples (fPDF) of all 63\,701 objects in the same tiles as the ALMA objects, from high (light red) to low (dark red). If XID+ was simply returning the prior, all the points would lie close to or along the red one-to-one line. This is evidently not the case for the ALMA objects, with extracted flux densities up to 1.5~dex away from their flux density priors, as well as the fPDF. The 350~$\mu$m and 500~$\mu$m results are similar.

\subsection{Comparisons of ALMA data to CIGALE SEDs}
The SPIRE data constrain the peak of the cold dust emission in a galaxy's SED and, if extracted correctly, any data in the long wavelength tail should lie on the SED. To test this, the SPIRE data from XID+, along with the data from the COSMOS2015 catalogue, were run through CIGALE to create the best fitting SED for all of the data. One object failed to converge, resulting in no output from CIGALE, and so had to be omitted. ALMA data were then compared to the best fitting SEDs.

\begin{figure}[t]
	\centering
	\includegraphics[width=0.43\textwidth]{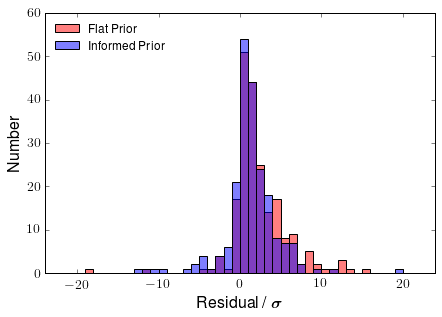}
	\caption{Distribution of the residuals between the measured ALMA flux densities and the flux densities inferred from the SEDs at the ALMA wavelengths expressed as a multiple of the error on the ALMA flux density ($\sigma$) for the flat prior (red) and the informed Gaussian prior (blue). The informed prior reduces the mean from 3.44~$\sigma$ to 1.83~$\sigma$, the standard deviation from 12.21~$\sigma$ to 7.95~$\sigma$ and the skew from 11.50 to 7.97.}
	\label{fig:sigma}
\end{figure}

For the 213 objects that successfully ran through CIGALE, the residuals between the measured ALMA flux densities and the ALMA flux densities inferred from the best fitting SEDs were calculated. These residuals ($\gamma$) are presented in Table \ref{table:sigma} and Fig. \ref{fig:sigma} as multiples of the errors on the ALMA flux densities ($\sigma$),  i.e. $\gamma$~=~residual~/~$\sigma$. A positive $\gamma$ means that the measured ALMA flux density is greater than the inferred flux density and the closer the value of $\gamma$ is to zero, the better XID+ had performed.

\begin{table}
	\caption{The absolute residuals (|$\gamma$|) between the measured ALMA flux densities and the fluxes of the best fitting SEDs at the ALMA wavelengths expressed as a multiple of the error on the ALMA flux density. The values in columns 2 and 3 are the number (percentage) of objects with a |$\gamma$| less than the value in column 1.}
	\label{table:sigma}
	\centering
	\begin{tabular}{l c c}
		\hline
		|$\gamma$| & Flat Prior & Informed Prior \\
		\hline
		1 & 77 (36.2\%) & 98 (46.0\%)\\
		2 & 121 (56.8\%) & 142 (66.7\%)\\
		3 & 146 (68.5\%) & 166 (77.9\%)\\
		4 & 160 (75.1\%) & 184 (86.4\%)\\
		5 & 177 (83.1\%)& 192 (90.1\%)\\
		\hline
	\end{tabular}
\end{table}

Figure \ref{fig:sigma} demonstrates the improvement that the informed prior gives over the flat prior. The informed Gaussian prior has a reduced spread around zero, a standard deviation of 7.95~$\sigma$ compared to 12.21~$\sigma$ for the flat prior, indicating that the informed prior is providing a better fit to the ALMA data than the flat prior. A reduced skew for the informed prior, with a skew of 7.97, with respect to the flat prior, skew of 11.50, and reduction in the mean of the distribution from 3.44~$\sigma$, for the flat prior, to 1.83~$\sigma$, for the informed prior, also demonstrates improved performance. With the reduced skew and lower mean, XID+ does not appear to under predict the SPIRE flux densities as often with the informed prior. Table \ref{table:sigma} shows the number of objects that have an absolute value of $\gamma$ (|$\gamma$|) below a certain threshold. The number of objects below each |$\gamma$| value is greater for the informed prior than it is for the flat prior, demonstrating that there is better agreement between the inferred ALMA flux densities of the best fitting SEDs and the measured data when using the informed prior than when using the flat prior. There is an increase of 27.3\% of the number of sources at |$\gamma$| < 1 when using the informed prior over the flat prior. This reduces to 8.5\% at |$\gamma$| < 5 and gives an average increase of 16.4\% of the number of sources across all five ranges. The maximum |$\gamma$| values follow the same trend with the flat prior giving a maximum |$\gamma$| of 167.304 compared to 95.180 for the informed prior. The minimum |$\gamma$| for the flat prior is that same as that of the informed prior.

\begin{figure}
	\centering
	\includegraphics[width=0.43\textwidth]{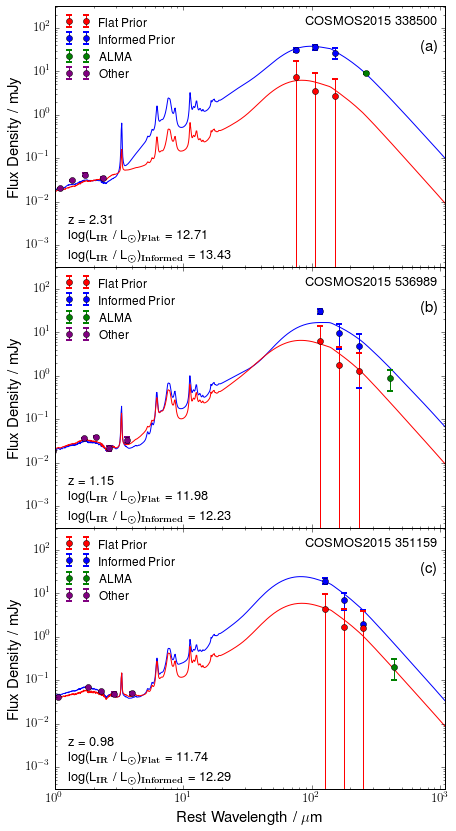}
	\caption{Example plots showing the best fitting SEDs (curves) for the extracted SPIRE flux densities (red and blue points) using the flat prior (red) and informed Gaussian prior (blue). The other multi-wavelength data from the COSMOS2015 catalogue (purple) are over plotted. Panel (a) is an example of Group A, illustrating how the informed prior can increase the agreement between the best fitting SED and the measured ALMA flux density (green). Panel (b) is for Group B, illustrating how the informed and flat priors can give equal agreement between the best fitting SED and the measured ALMA flux density. Panel (c) is for Group C, showing how the flat prior can occasionally give better agreement.}
	\label{fig:SED}
\end{figure}

The ALMA objects can be split into three groups: Group A, where the informed Gaussian prior provides a |$\gamma$| value more than 5\% smaller than the flat prior |$\gamma$|, Group B, where the informed prior |$\gamma$| is within 5\% of the flat prior |$\gamma$|, and Group C, where the informed |$\gamma$| is more than 5\% larger than the flat |$\gamma$|. Figure \ref{fig:SED}a provides an example of one of the 99 Group A objects. It can clearly be seen that the SED for the informed prior (blue line) is in much better agreement with the measured ALMA flux density (green point) than the SED for the flat prior (red line). There are 70 objects in Group B while Group C contains the remaining 44 objects, examples of which can be found in Fig. \ref{fig:SED}b and \ref{fig:SED}c respectively. With Group A being the largest group, it is evident that the informed prior produces more accurate results than the flat prior. If the objects with less than 2 ALMA signal-to-noise (see Sect. \ref{sec:data}) are removed, the number of objects in Groups A, B and C are 89, 51 and 35 respectively. Thus, the removal of these objects does not change the conclusion that, on average, the informed prior produces more accurate results.

\begin{figure}
	\centering
	\includegraphics[width=0.44\textwidth]{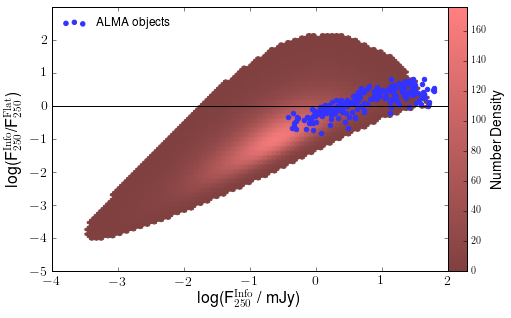}
	\caption{Plot of the ratio of the 250~$\mu$m flux densities extracted using the informed Gaussian prior (F$_{250}^{\mathrm{Info}}$) to those extracted with the flat prior (F$_{250}^{\mathrm{Flat}}$) against (F$_{250}^{\mathrm{Info}}$). The ALMA sources are in dark blue while the statistical average number density is from light red (high) to dark red (low).}
	\label{fig:flux}
\end{figure}

To see how the two XID+ priors compare with each other, Fig. \ref{fig:flux} shows the ratio between the flux density extracted using the informed prior (F$_{250}^{\mathrm{Info}}$) and flat prior (F$_{250}^{\mathrm{Flat}}$) against F$_{250}^{\mathrm{Info}}$. The ALMA sources, dark blue, show that as F$_{250}^{\mathrm{Info}}$ increases for the ALMA objects, the ratio F$_{250}^{\mathrm{Info}}$/F$_{250}^{\mathrm{Flat}}$ increases, becoming unity at F$_{250}^{\mathrm{Info}} \approx 3$~mJy, with smaller ratios below this flux density and higher ratios above. This implies that the flat prior over estimates flux densities for objects fainter than 3~mJy and under estimates flux densities for objects brighter than 3~mJy, with respect to the informed prior. This trend also appears for the fPDF, shown from high, light red, to low, dark red. There is an indication that the relation begins to flatten for high flux densities, which would be expected as both priors should perform equally well at higher flux densities. However, F$_{250}^{\mathrm{Info}}$/F$_{250}^{\mathrm{Flat}}$ being above unity is surprising. This may be the high source density affecting the flat prior results: the flat prior may be being "too democratic" as it assigns flux density to each object and so assigns too much to the faint objects and too little to the bright objects.

\begin{figure}
	\centering
	\includegraphics[width=0.43\textwidth]{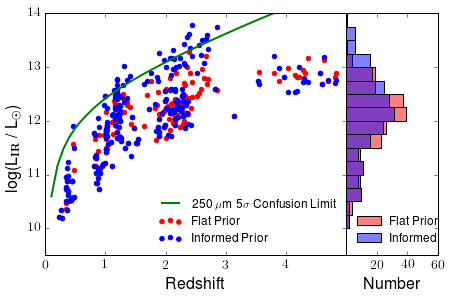}
	\caption{Scatter plot of the infrared (IR, 3~$\mu$m - 1100~$\mu$m) luminosity vs redshift for the results from XID+ with the flat prior (red points/bars) and the informed Gaussian prior (blue points/bars) with a histogram of the IR luminosities. The green line corresponds to the 250~$\mu$m 5$\sigma$ confusion limit.}
	\label{fig:IR_lum}
\end{figure}

Figure \ref{fig:IR_lum} illustrates how the difference between the flat and informed Gaussian XID+ priors affects the IR luminosity (L$_{\mathrm{IR}}$). The L$_{\mathrm{IR}}$ was calculated by integrating the best fitting SED between 3 and 1100~$\mu$m at rest. For the low ($z < 1$) and high ($z > 3$) redshift objects, the change in prior has little effect in L$_{\mathrm{IR}}$. However, for the intermediate redshift ($1 < z < 3$) objects, there appears to be an increase in L$_{\mathrm{IR}}$, although the number of objects in the sample is too small to draw any robust conclusions. As can be seen in the right panel of Fig. \ref{fig:IR_lum}, this results in a smoother distribution of luminosities across the sample presented. For comparison, the green line in Fig. \ref{fig:IR_lum} shows the 5$\sigma$ confusion limit for the 250~$\mu$m band, calculated by scaling the \citet{2008ApJ...682..985W} SED template to the 250~$\mu$m 5$\sigma$ confusion limit of 24.0~mJy and integrating between 3 and 1100 $\mu$m for a range of redshifts. Twenty of the objects with an informed prior, and one with a flat prior, do not fall below this confusion limit.

\subsection{Model Dependance}
To check if the difference between the informed prior and flat prior results is caused by our CIGALE model, we also fitted the data from the COSMOS2015 catalogue and the extracted SPIRE data for the 213 ALMA sources with MAGPHYS\footnote{\url{http://www.iap.fr/magphys/magphys/MAGPHYS.html}} \citep{2008MNRAS.388.1595D} as well as fitting modified black body (MBB) functions to the SPIRE data from the 213 sources. MAGPHYS is a SED modelling and fitting tool that uses different models from CIGALE, while MBBs are commonly used to model thermal dust emission \citep[e.g.][]{2013ApJ...772...77V, 2014PhR...541...45C, 2016MNRAS.461.1898W}. Here we use a MBB of the form
\begin{equation}
	S_{\nu} \propto \nu^{\beta} \times B_{\nu}(T)
\end{equation}
where B$_{\nu}(T)$ is the black body function for dust at temperature $T$. The emissivity, $\beta$ was fixed at 1.5 and $T$ was allowed to vary between 10K and 150K. With these results, we recreated Table \ref{table:sigma} for the MBB and MAGPHYS results as well as examined the populations of the three groups, A, B and C.

For the MBB, qualitatively the results are the same, although the division between the flat and informed priors is less wide. The results, shown in Table \ref{table:sigmaMbbMagphys}, show that the number of objects within certain |$\gamma$| thresholds is greater for the informed prior than the flat prior, supporting the idea that the results from XID+ using the informed prior are more reliable than the results with the flat prior. There is an average increase of 12.4\% of the number of sources across all five bins, which is less than the average over all five bins when comparing with CIGALE. Also, looking at the distribution of objects in the three groups, we find that Group A again has the most objects, with 126, while Groups B and C have fewer, with 37 and 51 objects respectively.
\begin{table}
	\caption{The absolute residuals (|$\gamma$|) between the measured ALMA flux densities and the fluxes of the best fitting MBB templates and MAGPHYS SEDs at the ALMA wavelengths expressed as a multiple of the error on the ALMA flux density. The values in columns 3 and 4 are the number (percentage) of objects with a |$\gamma$| less than the value in column 2.}
	\label{table:sigmaMbbMagphys}
	\centering
	\begin{tabular}{l c c c}
		\hline
		Model & |$\gamma$| & Flat Prior & Informed Prior \\
		\hline
		MBB & 1 & 62 (29.1\%) & 70 (32.9\%)\\
		& 2 & 106 (49.8\%) & 120 (56.3\%)\\
		& 3 & 134 (62.9\%) & 155 (72.8\%)\\
		& 4 & 152 (71.4\%) & 171 (80.3\%)\\
		& 5 & 168 (78.9\%) & 181 (85.0\%)\\
		\hline
		MAGPHYS & 1 & 72 (33.8\%) & 79 (37.1\%)\\
		& 2 & 119 (55.9\%) & 132 (62.0\%)\\
		& 3 & 138 (64.8\%) & 160 (75.1\%)\\
		& 4 & 156 (73.2\%) & 169 (79.3\%)\\
		& 5 & 171 (80.3\%) & 177 (83.1\%)\\
		\hline
	\end{tabular}
\end{table}

As with the results from the MBB, the results from MAGPHYS also have the same qualitative results as those using CIGALE. For MAGPHYS, the data from COSMOS2015 along with the extracted flux densities for the SPIRE bands were used and SEDs were fitted to these data. Similar to CIGALE, the flux densities at 870~$\mu$m and 1250~$\mu$m were extracted from the best fitting SEDs and compared to the ALMA data. The results are presented in Table \ref{table:sigmaMbbMagphys} and again show a greater number of objects within the values of |$\gamma$| examined for the informed prior with respect to the flat prior, with an average increase of 9.7\% of the number of sources across all the five |$\gamma$| ranges. As with the CIGALE and MBB results, Group A again contains the greatest number of objects, 108, followed by Group C, 65 objects, and Group B, 41 objects. As both the MAGPHYS and the MBB results are consistent with the CIGALE results, we believe that there is little model dependence in our conclusions.

\section{Conclusions} \label{sec:conc}
XID+ is one of the most advanced de-blending tools available. Here, we extend the flux density prior from as simple flat prior \citep{2017MNRAS.464..885H} and explore the idea of using supplementary data to provide a more physically motivated flux density prior. Using a rich set of multi-wavelength data, an informed Gaussian prior was introduced to XID+ and applied to 214 objects with ALMA detections in the COSMOS field. Using this method, it was shown that the agreement between the measured ALMA flux densities and the inferred flux densities at the ALMA wavelengths from the SEDs that best fit the XID+ extracted SPIRE flux densities, with CIGALE, improved by an average of 16.4\% of the number of objects in each of the five |$\gamma$| ranges examined in Sec. \ref{sec:discuss}. If all three models used for the ALMA comparison are considered, there is an average increase of 12.8\% demonstrating that, qualitatively, the results are not model dependent. However, the exact quantitative improvement seen is model dependent. This demonstrates that utilising supplementary data to provide a more physically motivated prior results in extracted flux density values that have improved agreement with data at other wavelengths.

\begin{acknowledgements}
We would like to thank the anonymous referee whose comments greatly improved this paper.
We would like to thank the Center for Information Technology of the University of Groningen for their support and for providing access to the Peregrine high performance computing cluster.
This research has made use of data from HerMES project (http://hermes.sussex.ac.uk/). HerMES is a Herschel Key Programme utilising Guaranteed Time from the SPIRE instrument team, ESAC scientists and a mission scientist.
The HerMES data were accessed through the \textit{Herschel} Database in Marseille (HeDaM - http://hedam.lam.fr) operated by CeSAM and hosted by the Laboratoire d'Astrophysique de Marseille.
HerMES DR4 was made possible through support of the Herschel Extragalactic Legacy Project, HELP (http://herschel.sussex.ac.uk).
The research leading to these results has received funding from the European Union Seventh Framework Programme FP7/2007-2013/ under grant agreement n\degr 607254. This publication reflects only the author's view and the European Union is not responsible for any use that may be made of the information contained therein.
\end{acknowledgements}

\bibliographystyle{aa}
\bibliography{Paper-AA-2016-30105}

\begin{thebibliography}{39}
\expandafter\ifx\csname natexlab\endcsname\relax\def\natexlab#1{#1}\fi

\bibitem[{{Bruzual} \& {Charlot}(2003)}]{2003MNRAS.344.1000B}
{Bruzual}, G. \& {Charlot}, S. 2003, \mnras, 344, 1000

\bibitem[{{Burgarella} {et~al.}(2013){Burgarella}, {Buat}, {Gruppioni},
  {Cucciati}, {Heinis}, {Berta}, {B{\'e}thermin}, {Bock}, {Cooray}, {Dunlop},
  {Farrah}, {Franceschini}, {Le Floc'h}, {Lutz}, {Magnelli}, {Nordon},
  {Oliver}, {Page}, {Popesso}, {Pozzi}, {Riguccini}, {Vaccari}, \&
  {Viero}}]{2013A&A...554A..70B}
{Burgarella}, D., {Buat}, V., {Gruppioni}, C., {et~al.} 2013, \aap, 554, A70

\bibitem[{{Calzetti} {et~al.}(2000){Calzetti}, {Armus}, {Bohlin}, {Kinney},
  {Koornneef}, \& {Storchi-Bergmann}}]{2000ApJ...533..682C}
{Calzetti}, D., {Armus}, L., {Bohlin}, R.~C., {et~al.} 2000, \apj, 533, 682

\bibitem[{Carpenter {et~al.}(2017)Carpenter, Gelman, Hoffman, Lee, Goodrich,
  Betancourt, Brubaker, Guo, Li, \& Riddell}]{JSSv076i01}
Carpenter, B., Gelman, A., Hoffman, M., {et~al.} 2017, Journal of Statistical
  Software, 76, 1

\bibitem[{{Casey} {et~al.}(2014){Casey}, {Narayanan}, \&
  {Cooray}}]{2014PhR...541...45C}
{Casey}, C.~M., {Narayanan}, D., \& {Cooray}, A. 2014, \physrep, 541, 45

\bibitem[{{Chabrier}(2003)}]{2003PASP..115..763C}
{Chabrier}, G. 2003, \pasp, 115, 763

\bibitem[{{Ciesla} {et~al.}(2015){Ciesla}, {Charmandaris}, {Georgakakis},
  {Bernhard}, {Mitchell}, {Buat}, {Elbaz}, {LeFloc'h}, {Lacey}, {Magdis}, \&
  {Xilouris}}]{2015AA...576A..10C}
{Ciesla}, L., {Charmandaris}, V., {Georgakakis}, A., {et~al.} 2015, \aap, 576,
  A10

\bibitem[{{da Cunha} {et~al.}(2008){da Cunha}, {Charlot}, \&
  {Elbaz}}]{2008MNRAS.388.1595D}
{da Cunha}, E., {Charlot}, S., \& {Elbaz}, D. 2008, \mnras, 388, 1595

\bibitem[{{Dole} {et~al.}(2006){Dole}, {Lagache}, {Puget}, {Caputi},
  {Fern{\'a}ndez-Conde}, {Le Floc'h}, {Papovich}, {P{\'e}rez-Gonz{\'a}lez},
  {Rieke}, \& {Blaylock}}]{2006AA...451..417D}
{Dole}, H., {Lagache}, G., {Puget}, J.-L., {et~al.} 2006, \aap, 451, 417

\bibitem[{{Dole} {et~al.}(2004){Dole}, {Rieke}, {Lagache}, {Puget},
  {Alonso-Herrero}, {Bai}, {Blaylock}, {Egami}, {Engelbracht}, {Gordon},
  {Hines}, {Kelly}, {Le Floc'h}, {Misselt}, {Morrison}, {Muzerolle},
  {Papovich}, {P{\'e}rez-Gonz{\'a}lez}, {Rieke}, {Rigby}, {Neugebauer},
  {Stansberry}, {Su}, {Young}, {Beichman}, \& {Richards}}]{2004ApJS..154...93D}
{Dole}, H., {Rieke}, G.~H., {Lagache}, G., {et~al.} 2004, \apjs, 154, 93

\bibitem[{{Draine}(2003)}]{2003ARAA..41..241D}
{Draine}, B.~T. 2003, \araa, 41, 241

\bibitem[{{Draine} {et~al.}(2014){Draine}, {Aniano}, {Krause}, {Groves},
  {Sandstrom}, {Braun}, {Leroy}, {Klaas}, {Linz}, {Rix}, {Schinnerer},
  {Schmiedeke}, \& {Walter}}]{2014ApJ...780..172D}
{Draine}, B.~T., {Aniano}, G., {Krause}, O., {et~al.} 2014, \apj, 780, 172

\bibitem[{{Draine} \& {Li}(2007)}]{2007ApJ...657..810D}
{Draine}, B.~T. \& {Li}, A. 2007, \apj, 657, 810

\bibitem[{{Fritz} {et~al.}(2006){Fritz}, {Franceschini}, \&
  {Hatziminaoglou}}]{2006MNRAS.366..767F}
{Fritz}, J., {Franceschini}, A., \& {Hatziminaoglou}, E. 2006, \mnras, 366, 767

\bibitem[{{G{\'o}rski} {et~al.}(2005){G{\'o}rski}, {Hivon}, {Banday},
  {Wandelt}, {Hansen}, {Reinecke}, \& {Bartelmann}}]{2005ApJ...622..759G}
{G{\'o}rski}, K.~M., {Hivon}, E., {Banday}, A.~J., {et~al.} 2005, \apj, 622,
  759

\bibitem[{{Griffin} {et~al.}(2010){Griffin}, {Abergel}, {Abreu}, {Ade},
  {Andr{\'e}}, {Augueres}, {Babbedge}, {Bae}, {Baillie}, {Baluteau}, {Barlow},
  {Bendo}, {Benielli}, {Bock}, {Bonhomme}, {Brisbin}, {Brockley-Blatt},
  {Caldwell}, {Cara}, {Castro-Rodriguez}, {Cerulli}, {Chanial}, {Chen},
  {Clark}, {Clements}, {Clerc}, {Coker}, {Communal}, {Conversi}, {Cox},
  {Crumb}, {Cunningham}, {Daly}, {Davis}, {de Antoni}, {Delderfield}, {Devin},
  {di Giorgio}, {Didschuns}, {Dohlen}, {Donati}, {Dowell}, {Dowell}, {Duband},
  {Dumaye}, {Emery}, {Ferlet}, {Ferrand}, {Fontignie}, {Fox}, {Franceschini},
  {Frerking}, {Fulton}, {Garcia}, {Gastaud}, {Gear}, {Glenn}, {Goizel},
  {Griffin}, {Grundy}, {Guest}, {Guillemet}, {Hargrave}, {Harwit}, {Hastings},
  {Hatziminaoglou}, {Herman}, {Hinde}, {Hristov}, {Huang}, {Imhof}, {Isaak},
  {Israelsson}, {Ivison}, {Jennings}, {Kiernan}, {King}, {Lange}, {Latter},
  {Laurent}, {Laurent}, {Leeks}, {Lellouch}, {Levenson}, {Li}, {Li},
  {Lilienthal}, {Lim}, {Liu}, {Lu}, {Madden}, {Mainetti}, {Marliani}, {McKay},
  {Mercier}, {Molinari}, {Morris}, {Moseley}, {Mulder}, {Mur}, {Naylor},
  {Nguyen}, {O'Halloran}, {Oliver}, {Olofsson}, {Olofsson}, {Orfei}, {Page},
  {Pain}, {Panuzzo}, {Papageorgiou}, {Parks}, {Parr-Burman}, {Pearce},
  {Pearson}, {P{\'e}rez-Fournon}, {Pinsard}, {Pisano}, {Podosek}, {Pohlen},
  {Polehampton}, {Pouliquen}, {Rigopoulou}, {Rizzo}, {Roseboom}, {Roussel},
  {Rowan-Robinson}, {Rownd}, {Saraceno}, {Sauvage}, {Savage}, {Savini},
  {Sawyer}, {Scharmberg}, {Schmitt}, {Schneider}, {Schulz}, {Schwartz},
  {Shafer}, {Shupe}, {Sibthorpe}, {Sidher}, {Smith}, {Smith}, {Smith},
  {Spencer}, {Stobie}, {Sudiwala}, {Sukhatme}, {Surace}, {Stevens}, {Swinyard},
  {Trichas}, {Tourette}, {Triou}, {Tseng}, {Tucker}, {Turner}, {Vaccari},
  {Valtchanov}, {Vigroux}, {Virique}, {Voellmer}, {Walker}, {Ward}, {Waskett},
  {Weilert}, {Wesson}, {White}, {Whitehouse}, {Wilson}, {Winter}, {Woodcraft},
  {Wright}, {Xu}, {Zavagno}, {Zemcov}, {Zhang}, \&
  {Zonca}}]{2010AA...518L...3G}
{Griffin}, M.~J., {Abergel}, A., {Abreu}, A., {et~al.} 2010, \aap, 518, L3

\bibitem[{{Hurley} {et~al.}(2017){Hurley}, {Oliver}, {Betancourt}, {Clarke},
  {Cowley}, {Duivenvoorden}, {Farrah}, {Griffin}, {Lacey}, {Le Floc'h},
  {Papadopoulos}, {Sargent}, {Scudder}, {Vaccari}, {Valtchanov}, \&
  {Wang}}]{2017MNRAS.464..885H}
{Hurley}, P.~D., {Oliver}, S., {Betancourt}, M., {et~al.} 2017, \mnras, 464,
  885

\bibitem[{{Komatsu} {et~al.}(2011){Komatsu}, {Smith}, {Dunkley}, {Bennett},
  {Gold}, {Hinshaw}, {Jarosik}, {Larson}, {Nolta}, {Page}, {Spergel},
  {Halpern}, {Hill}, {Kogut}, {Limon}, {Meyer}, {Odegard}, {Tucker}, {Weiland},
  {Wollack}, \& {Wright}}]{2011ApJS..192...18K}
{Komatsu}, E., {Smith}, K.~M., {Dunkley}, J., {et~al.} 2011, \apjs, 192, 18

\bibitem[{{Laigle} {et~al.}(2016){Laigle}, {McCracken}, {Ilbert}, {Hsieh},
  {Davidzon}, {Capak}, {Hasinger}, {Silverman}, {Pichon}, {Coupon}, {Aussel},
  {Le Borgne}, {Caputi}, {Cassata}, {Chang}, {Civano}, {Dunlop}, {Fynbo},
  {Kartaltepe}, {Koekemoer}, {Le F{\`e}vre}, 
  {Lilly}, {Lin}, {Marchesi}, {Milvang-Jensen}, {Salvato}, {Sanders},
  {Scoville}, {Smolcic}, {Stockmann}, {Taniguchi}, {Tasca}, {Toft}, {Vaccari},
  \& {Zabl}}]{2016ApJS..224...24L}
{Laigle}, C., {McCracken}, H.~J., {Ilbert}, O., {et~al.} 2016, \apjs, 224, 24

\bibitem[{{Larson} {et~al.}(2011){Larson}, {Dunkley}, {Hinshaw}, {Komatsu},
  {Nolta}, {Bennett}, {Gold}, {Halpern}, {Hill}, {Jarosik}, {Kogut}, {Limon},
  {Meyer}, {Odegard}, {Page}, {Smith}, {Spergel}, {Tucker}, {Weiland},
  {Wollack}, \& {Wright}}]{2011ApJS..192...16L}
{Larson}, D., {Dunkley}, J., {Hinshaw}, G., {et~al.} 2011, \apjs, 192, 16

\bibitem[{{Merlin} {et~al.}(2015){Merlin}, {Fontana}, {Ferguson}, {Dunlop},
  {Elbaz}, {Bourne}, {Bruce}, {Buitrago}, {Castellano}, {Schreiber}, {Grazian},
  {McLure}, {Okumura}, {Shu}, {Wang}, {Amor{\'{\i}}n}, {Boutsia}, {Cappelluti},
  {Comastri}, {Derriere}, {Faber}, \& {Santini}}]{2015AA...582A..15M}
{Merlin}, E., {Fontana}, A., {Ferguson}, H.~C., {et~al.} 2015, \aap, 582, A15

\bibitem[{{Mitchell} {et~al.}(2013){Mitchell}, {Lacey}, {Baugh}, \&
  {Cole}}]{2013MNRAS.435...87M}
{Mitchell}, P.~D., {Lacey}, C.~G., {Baugh}, C.~M., \& {Cole}, S. 2013, \mnras,
  435, 87

\bibitem[{{Nguyen} {et~al.}(2010){Nguyen}, {Schulz}, {Levenson}, {Amblard},
  {Arumugam}, {Aussel}, {Babbedge}, {Blain}, {Bock}, {Boselli}, {Buat},
  {Castro-Rodriguez}, {Cava}, {Chanial}, {Chapin}, {Clements}, {Conley},
  {Conversi}, {Cooray}, {Dowell}, {Dwek}, {Eales}, {Elbaz}, {Fox},
  {Franceschini}, {Gear}, {Glenn}, {Griffin}, {Halpern}, {Hatziminaoglou},
  {Ibar}, {Isaak}, {Ivison}, {Lagache}, {Lu}, {Madden}, {Maffei}, {Mainetti},
  {Marchetti}, {Marsden}, {Marshall}, {O'Halloran}, {Oliver}, {Omont}, {Page},
  {Panuzzo}, {Papageorgiou}, {Pearson}, {Perez Fournon}, {Pohlen}, {Rangwala},
  {Rigopoulou}, {Rizzo}, {Roseboom}, {Rowan-Robinson}, {Scott}, {Seymour},
  {Shupe}, {Smith}, {Stevens}, {Symeonidis}, {Trichas}, {Tugwell}, {Vaccari},
  {Valtchanov}, {Vigroux}, {Wang}, {Ward}, {Wiebe}, {Wright}, {Xu}, \&
  {Zemcov}}]{2010A&A...518L...5N}
{Nguyen}, H.~T., {Schulz}, B., {Levenson}, L., {et~al.} 2010, \aap, 518, L5

\bibitem[{{Noll} {et~al.}(2009){Noll}, {Burgarella}, {Giovannoli}, {Buat},
  {Marcillac}, \& {Mu{\~n}oz-Mateos}}]{2009AA...507.1793N}
{Noll}, S., {Burgarella}, D., {Giovannoli}, E., {et~al.} 2009, \aap, 507, 1793

\bibitem[{{Oliver} {et~al.}(2012){Oliver}, {Bock}, {Altieri}, {Amblard},
  {Arumugam}, {Aussel}, {Babbedge}, {Beelen}, {B{\'e}thermin}, {Blain},
  {Boselli}, {Bridge}, {Brisbin}, {Buat}, {Burgarella},
  {Castro-Rodr{\'{\i}}guez}, {Cava}, {Chanial}, {Cirasuolo}, {Clements},
  {Conley}, {Conversi}, {Cooray}, {Dowell}, {Dubois}, {Dwek}, {Dye}, {Eales},
  {Elbaz}, {Farrah}, {Feltre}, {Ferrero}, {Fiolet}, {Fox}, {Franceschini},
  {Gear}, {Giovannoli}, {Glenn}, {Gong}, {Gonz{\'a}lez Solares}, {Griffin},
  {Halpern}, {Harwit}, {Hatziminaoglou}, {Heinis}, {Hurley}, {Hwang}, {Hyde},
  {Ibar}, {Ilbert}, {Isaak}, {Ivison}, {Lagache}, {Le Floc'h}, {Levenson},
  {Faro}, {Lu}, {Madden}, {Maffei}, {Magdis}, {Mainetti}, {Marchetti},
  {Marsden}, {Marshall}, {Mortier}, {Nguyen}, {O'Halloran}, {Omont}, {Page},
  {Panuzzo}, {Papageorgiou}, {Patel}, {Pearson}, {P{\'e}rez-Fournon}, {Pohlen},
  {Rawlings}, {Raymond}, {Rigopoulou}, {Riguccini}, {Rizzo}, {Rodighiero},
  {Roseboom}, {Rowan-Robinson}, {S{\'a}nchez Portal}, {Schulz}, {Scott},
  {Seymour}, {Shupe}, {Smith}, {Stevens}, {Symeonidis}, {Trichas}, {Tugwell},
  {Vaccari}, {Valtchanov}, {Vieira}, {Viero}, {Vigroux}, {Wang}, {Ward},
  {Wardlow}, {Wright}, {Xu}, \& {Zemcov}}]{2012MNRAS.424.1614O}
{Oliver}, S.~J., {Bock}, J., {Altieri}, B., {et~al.} 2012, \mnras, 424, 1614

\bibitem[{{Pilbratt} {et~al.}(2010){Pilbratt}, {Riedinger}, {Passvogel},
  {Crone}, {Doyle}, {Gageur}, {Heras}, {Jewell}, {Metcalfe}, {Ott}, \&
  {Schmidt}}]{2010AA...518L...1P}
{Pilbratt}, G.~L., {Riedinger}, J.~R., {Passvogel}, T., {et~al.} 2010, \aap,
  518, L1

\bibitem[{{Puglisi} {et~al.}(2016){Puglisi}, {Rodighiero}, {Franceschini},
  {Talia}, {Cimatti}, {Baronchelli}, {Daddi}, {Renzini}, {Schawinski},
  {Mancini}, {Silverman}, {Gruppioni}, {Lutz}, {Berta}, \&
  {Oliver}}]{2016AA...586A..83P}
{Puglisi}, A., {Rodighiero}, G., {Franceschini}, A., {et~al.} 2016, \aap, 586,
  A83

\bibitem[{{Rieke} {et~al.}(2004){Rieke}, {Young}, {Engelbracht}, {Kelly},
  {Low}, {Haller}, {Beeman}, {Gordon}, {Stansberry}, {Misselt}, {Cadien},
  {Morrison}, {Rivlis}, {Latter}, {Noriega-Crespo}, {Padgett}, {Stapelfeldt},
  {Hines}, {Egami}, {Muzerolle}, {Alonso-Herrero}, {Blaylock}, {Dole}, {Hinz},
  {Le Floc'h}, {Papovich}, {P{\'e}rez-Gonz{\'a}lez}, {Smith}, {Su}, {Bennett},
  {Frayer}, {Henderson}, {Lu}, {Masci}, {Pesenson}, {Rebull}, {Rho}, {Keene},
  {Stolovy}, {Wachter}, {Wheaton}, {Werner}, \&
  {Richards}}]{2004ApJS..154...25R}
{Rieke}, G.~H., {Young}, E.~T., {Engelbracht}, C.~W., {et~al.} 2004, \apjs,
  154, 25

\bibitem[{{Roseboom} {et~al.}(2012){Roseboom}, {Ivison}, {Greve}, {Amblard},
  {Arumugam}, {Auld}, {Aussel}, {Bethermin}, {Blain}, {Bock}, {Boselli},
  {Brisbin}, {Buat}, {Burgarella}, {Castro-Rodr{\'{\i}}guez}, {Cava},
  {Chanial}, {Chapin}, {Chapman}, {Clements}, {Conley}, {Conversi}, {Cooray},
  {Dowell}, {Dunlop}, {Dwek}, {Eales}, {Elbaz}, {Farrah}, {Franceschini},
  {Glenn}, {Griffin}, {Halpern}, {Hatziminaoglou}, {Ibar}, {Isaak}, {Lagache},
  {Levenson}, {Lu}, {Madden}, {Maffei}, {Mainetti}, {Marchetti}, {Marsden},
  {Morrison}, {Mortier}, {Nguyen}, {O'Halloran}, {Oliver}, {Omont}, {Page},
  {Panuzzo}, {Papageorgiou}, {Pearson}, {P{\'e}rez-Fournon}, {Pohlen},
  {Rawlings}, {Raymond}, {Rigopoulou}, {Rizzo}, {Rodighiero}, {Rowan-Robinson},
  {Schulz}, {Scott}, {Seymour}, {Shupe}, {Smith}, {Stevens}, {Symeonidis},
  {Trichas}, {Tugwell}, {Vaccari}, {Valtchanov}, {Vieira}, {Viero}, {Vigroux},
  {Wardlow}, {Wang}, {Wright}, {Xu}, \& {Zemcov}}]{2012MNRAS.419.2758R}
{Roseboom}, I.~G., {Ivison}, R.~J., {Greve}, T.~R., {et~al.} 2012, \mnras, 419,
  2758

\bibitem[{{Roseboom} {et~al.}(2010){Roseboom}, {Oliver}, {Kunz}, {Altieri},
  {Amblard}, {Arumugam}, {Auld}, {Aussel}, {Babbedge}, {B{\'e}thermin},
  {Blain}, {Bock}, {Boselli}, {Brisbin}, {Buat}, {Burgarella},
  {Castro-Rodr{\'{\i}}guez}, {Cava}, {Chanial}, {Chapin}, {Clements}, {Conley},
  {Conversi}, {Cooray}, {Dowell}, {Dwek}, {Dye}, {Eales}, {Elbaz}, {Farrah},
  {Fox}, {Franceschini}, {Gear}, {Glenn}, {Solares}, {Griffin}, {Halpern},
  {Harwit}, {Hatziminaoglou}, {Huang}, {Ibar}, {Isaak}, {Ivison}, {Lagache},
  {Levenson}, {Lu}, {Madden}, {Maffei}, {Mainetti}, {Marchetti}, {Marsden},
  {Mortier}, {Nguyen}, {O'Halloran}, {Omont}, {Page}, {Panuzzo},
  {Papageorgiou}, {Patel}, {Pearson}, {P{\'e}rez-Fournon}, {Pohlen},
  {Rawlings}, {Raymond}, {Rigopoulou}, {Rizzo}, {Rowan-Robinson}, {Portal},
  {Schulz}, {Scott}, {Seymour}, {Shupe}, {Smith}, {Stevens}, {Symeonidis},
  {Trichas}, {Tugwell}, {Vaccari}, {Valtchanov}, {Vieira}, {Vigroux}, {Wang},
  {Ward}, {Wright}, {Xu}, \& {Zemcov}}]{2010MNRAS.409...48R}
{Roseboom}, I.~G., {Oliver}, S.~J., {Kunz}, M., {et~al.} 2010, \mnras, 409, 48

\bibitem[{{Scoville} {et~al.}(2007){Scoville}, {Aussel}, {Brusa}, {Capak},
  {Carollo}, {Elvis}, {Giavalisco}, {Guzzo}, {Hasinger}, {Impey}, {Kneib},
  {LeFevre}, {Lilly}, {Mobasher}, {Renzini}, {Rich}, {Sanders}, {Schinnerer},
  {Schminovich}, {Shopbell}, {Taniguchi}, \& {Tyson}}]{2007ApJS..172....1S}
{Scoville}, N., {Aussel}, H., {Brusa}, M., {et~al.} 2007, \apjs, 172, 1

\bibitem[{{Scoville} {et~al.}(2014){Scoville}, {Aussel}, {Sheth}, {Scott},
  {Sanders}, {Ivison}, {Pope}, {Capak}, {Vanden Bout}, {Manohar}, {Kartaltepe},
  {Robertson}, \& {Lilly}}]{2014ApJ...783...84S}
{Scoville}, N., {Aussel}, H., {Sheth}, K., {et~al.} 2014, \apj, 783, 84

\bibitem[{{Scoville} {et~al.}(2016){Scoville}, {Sheth}, {Aussel}, {Vanden
  Bout}, {Capak}, {Bongiorno}, {Casey}, {Murchikova}, {Koda},
  {{\'A}lvarez-M{\'a}rquez}, {Lee}, {Laigle}, {McCracken}, {Ilbert}, {Pope},
  {Sanders}, {Chu}, {Toft}, {Ivison}, \& {Manohar}}]{2016ApJ...820...83S}
{Scoville}, N., {Sheth}, K., {Aussel}, H., {et~al.} 2016, \apj, 820, 83

\bibitem[{{Scudder} {et~al.}(2016){Scudder}, {Oliver}, {Hurley}, {Griffin},
  {Sargent}, {Scott}, {Wang}, \& {Wardlow}}]{2016MNRAS.460.1119S}
{Scudder}, J.~M., {Oliver}, S., {Hurley}, P.~D., {et~al.} 2016, \mnras, 460,
  1119

\bibitem[{{Serra} {et~al.}(2011){Serra}, {Amblard}, {Temi}, {Burgarella},
  {Giovannoli}, {Buat}, {Noll}, \& {Im}}]{2011ApJ...740...22S}
{Serra}, P., {Amblard}, A., {Temi}, P., {et~al.} 2011, \apj, 740, 22

\bibitem[{{Viero} {et~al.}(2013){Viero}, {Wang}, {Zemcov}, {Addison},
  {Amblard}, {Arumugam}, {Aussel}, {B{\'e}thermin}, {Bock}, {Boselli}, {Buat},
  {Burgarella}, {Casey}, {Clements}, {Conley}, {Conversi}, {Cooray}, {De
  Zotti}, {Dowell}, {Farrah}, {Franceschini}, {Glenn}, {Griffin},
  {Hatziminaoglou}, {Heinis}, {Ibar}, {Ivison}, {Lagache}, {Levenson},
  {Marchetti}, {Marsden}, {Nguyen}, {O'Halloran}, {Oliver}, {Omont}, {Page},
  {Papageorgiou}, {Pearson}, {P{\'e}rez-Fournon}, {Pohlen}, {Rigopoulou},
  {Roseboom}, {Rowan-Robinson}, {Schulz}, {Scott}, {Seymour}, {Shupe}, {Smith},
  {Symeonidis}, {Vaccari}, {Valtchanov}, {Vieira}, {Wardlow}, \&
  {Xu}}]{2013ApJ...772...77V}
{Viero}, M.~P., {Wang}, L., {Zemcov}, M., {et~al.} 2013, \apj, 772, 77

\bibitem[{{Wang} {et~al.}(2016){Wang}, {Norberg}, {Gunawardhana}, {Heinis},
  {Baldry}, {Bland-Hawthorn}, {Bourne}, {Brough}, {Brown}, {Cluver}, {Cooray},
  {da Cunha}, {Driver}, {Dunne}, {Dye}, {Eales}, {Grootes}, {Holwerda},
  {Hopkins}, {Ibar}, {Ivison}, {Lacey}, {Lara-Lopez}, {Loveday}, {Maddox},
  {Micha{\l}owski}, {Oteo}, {Owers}, {Popescu}, {Smith}, {Taylor}, {Tuffs}, \&
  {van der Werf}}]{2016MNRAS.461.1898W}
{Wang}, L., {Norberg}, P., {Gunawardhana}, M.~L.~P., {et~al.} 2016, \mnras,
  461, 1898

\bibitem[{{Wang} {et~al.}(2014){Wang}, {Viero}, {Clarke}, {Bock}, {Buat},
  {Conley}, {Farrah}, {Guo}, {Heinis}, {Magdis}, {Marchetti}, {Marsden},
  {Norberg}, {Oliver}, {Page}, {Roehlly}, {Roseboom}, {Schulz}, {Smith},
  {Vaccari}, \& {Zemcov}}]{2014MNRAS.444.2870W}
{Wang}, L., {Viero}, M., {Clarke}, C., {et~al.} 2014, \mnras, 444, 2870

\bibitem[{{Wuyts} {et~al.}(2008){Wuyts}, {Labb{\'e}}, {F{\"o}rster Schreiber},
  {Franx}, {Rudnick}, {Brammer}, \& {van Dokkum}}]{2008ApJ...682..985W}
{Wuyts}, S., {Labb{\'e}}, I., {F{\"o}rster Schreiber}, N.~M., {et~al.} 2008,
  \apj, 682, 985

\end{thebibliography}


\begin{thebibliography}{}
\expandafter\ifx\csname natexlab\endcsname\relax\def\natexlab#1{#1}\fi

\bibitem[{{Zuckerman} \& {Palmer}(1974)}]{1974ARA&A..12..279Z}
{Zuckerman}, B., \& {Palmer}, P. 1974, \araa, 12, 279

\end{thebibliography}

\begin{appendix}
\section{Homogeneity of COSMOS}\label{app:noise}
\begin{figure}
	\centering
	\includegraphics[width=0.4\textwidth]{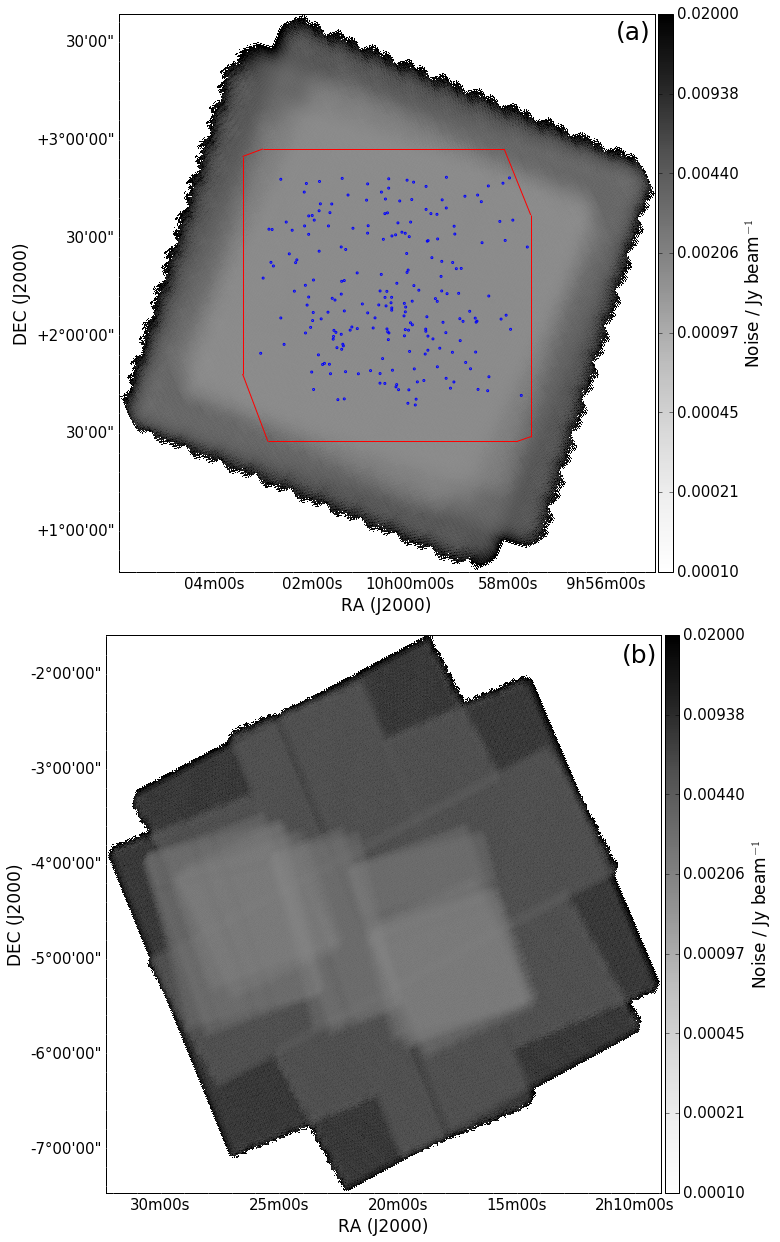}
	\caption{The noise map for the SPIRE 250~$\mu$m band in the (a) COSMOS field (2 deg$^{2}$) and (b) XMM-LSS field (11 deg$^{2}$) from the HerMES DR4 \citep{2012MNRAS.424.1614O} and downloaded from HeDAM. For the COSMOS field, the outline of the study area is shown in red, while the ALMA sources are shown in blue.}
	\label{fig:noise}
\end{figure}

Figure \ref{fig:noise}a shows the noise map for the 2 square degree 250~$\mu$m SPIRE band in the COSMOS field with the area used for this study outlined in red and the position of the ALMA sources shown with blue points. As can be seen, the noise level within this area is approximately flat. For comparison, the noise map for the 250~$\mu$m SPIRE band in the 11 square degree XMM-LSS field is shown in Fig. \ref{fig:noise}b, which is much less homogenous. The noise maps for the 350~$\mu$m and 500~$\mu$m bands are very similar.

\section{CIGALE Parameters}\label{app:parameters}
In this work, a double exponential SFH was used over the more commonly used exponentially declining or delayed exponential SFH as these two SFHs did not appear to reproduce the expected starburst population in the star formation rate vs. stellar mass plane. The e-folding time of the two stellar populations (old and young) in the SFH was roughly matched to that of \citet{2013MNRAS.435...87M}. As \citet{2013MNRAS.435...87M} used a single declining exponential SFH, the e-folding times were split with the burst population taking values of 9~Gyr and above and the main population taking values of less than 9~Gyr. For the ages of the main population, the values were sampled between 1 and 13~Gyr with a log distribution. The mass fraction of the burst population follows \citet{2015AA...576A..10C} along with the age of the young stellar population, which also had a lower age of 0.001~Gyr added. The \citet{2003MNRAS.344.1000B} stellar population model was used with a \citet{2003PASP..115..763C} initial mass function. As this study was not to explore the metallicity of galaxies, it was decided to leave the metallically at solar.

The dust colour excess (E(B-V)$^{\star}$) is often measured to be $\lesssim$~0.6, with a higher number of objects with lower excesses \citep[e.g.][]{2000ApJ...533..682C}, so this region has been sampled for the SED models. However, it is possible that the E(B-V)$^{\star}$ could be larger than this so values of 1.1 and 2.0 are also included. For the E(B-V)$^{\star}_{\mathrm{old}}$ reduction factor, it has been shown that a value of 0.58 is needed for the local universe \citep{2016AA...586A..83P} instead of the \citet{2000ApJ...533..682C} value of 0.44. A compromise of 0.5 was therefore used.

For the dust emission, the polycyclic aromatic hydrocarbon (PAH) fraction had an increase in range around the default 2.5 so more fractions could be sampled while keeping the number of models reasonable. The minimum scaling factor of the radiation field was similarly given a range to sample with an increase in the smallest value from 1.0 to 5.0. The illuminated fraction was reduced to 0.02, following \citet{2015AA...576A..10C}.

The parameters in the AGN model were matched to those used by \citet{2015AA...576A..10C}, who undertook a detailed study of AGN host galaxy emission using CIGALE. The number of choices of frac$_{AGN}$ was reduced from 14 to 9, while still covering the same range, to reduce the number of models created by CIGALE to decrease runtime.

A list of parameters, where they differ from default, can be found in Table \ref{table:params}.1.

\begin{table*}\label{table:params}
	\caption{Parameters used for the various properties in the CIGALE model SEDs where they different from the default values. All ages and times are in Gyr.}
	\label{table:parameters}
	\centering
	\begin{tabular}{l c c}
		\hline\hline
		Parameter & Value & Description\\
		\hline
		\multicolumn{3}{c}{Star Formation History} \\
		\hline
		& & \\
		$\tau_{main}$ & 1.0, 3.0, 5.0 & e-folding time (main)\\
		$\tau_{burst}$ & 9.0, 13.0 & e-folding time (burst) \\
		$f_{burst}$ & 0.001, 0.01, 0.1, 0.2 & Burst mass fraction\\
		age & 1.000, 1.329, 1.768, 2.351, 3.126, & Population age (main)\\
		& 4.157, 5.528, 7.352, 9.776, 13.000 & \\
		burst age & 0.001, 0.010, 0.030, 0.100, 0.300 & Population age (burst)\\
		& & \\
		\hline
		\multicolumn{3}{c}{Stellar Emission}\\
		\hline
		& & \\
		$Z$ & 0.02 & Metallicity (0.02 is Solar)\\
		& & \\
		\hline
		\multicolumn{3}{c}{Dust Attenuation}\\
		\hline
		& & \\
		$E(B-V)^{*}_{young}$ & 0.1, 0.18, 0.33, 0.6, 1.1, 2.0 & E(B-V)* for young population\\
		$E(B-V)^{*}_{old}$ & 0.5 & Reduction factor in E(B-V)* for old population\\
		& & \\
		\hline
		\multicolumn{3}{c}{Dust Emission}\\
		\hline
		& & \\
		$q_{PAH}$ & 1.12, 2.50, 3.19 & Mass fraction of PAH \\
		$U_{min}$ & 5.0, 10.0, 25.0 & Minimum scaling factor of the radiation field intensity\\
		$\gamma$ & 0.02 & Illuminated fraction\\
		& & \\
		\hline
		\multicolumn{3}{c}{AGN Emission}\\
		\hline
		& & \\
		$\tau$ & 1.0, 6.0 & Optical depth at 9.7~$\mu$m\\
		$\gamma$ & 0.0 & $\gamma$ coefficient for the gas density function of the torus\tablefootmark{a}\\
		$\psi$ & 0.001\degr, 89.990\degr & Angle between equatorial axis and line of sight\\
		$frac_{AGN}$ & 0.0, 0.05, 0.1, 0.2, 0.3, 0.4, 0.5 & AGN fraction\\
		 & 0.6, 0.7 &\\
		\hline
	\end{tabular}
	\tablefoot{
		\tablefoottext{a}{Density function of the torus can be found in \citet{2006MNRAS.366..767F} as Equation 3.}
	}
\end{table*}

\end{appendix}

\end{document}